\documentstyle[11pt]{article}

\def\AFOUR{%
\setlength{\textheight}{9.0in}%
\setlength{\textwidth}{5.75in}%
\setlength{\topmargin}{-0.375in}%
\hoffset=-.5in%
\renewcommand{\baselinestretch}{1.17}%
\setlength{\parskip}{6pt plus 2pt}%
}

\AFOUR                                           

\expandafter\ifx\csname amssym..def\endcsname\relax \else\endinput\fi
\expandafter\edef\csname amssym.def\endcsname{%
       \catcode`\noexpand\@=\the\catcode`\@\space}
\catcode`\@=11
\def\undefine#1{\let#1\undefined}
\def\newsymbol#1#2#3#4#5{\let\next@\relax
 \ifnum#2=\@ne\let\next@\msafam@\else
 \ifnum#2=\tw@\let\next@\msbfam@\fi\fi
 \mathchardef#1="#3\next@#4#5}
\def\mathhexbox@#1#2#3{\relax
 \ifmmode\mathpalette{}{\m@th\mathchar"#1#2#3}%
 \else\leavevmode\hbox{$\m@th\mathchar"#1#2#3$}\fi}
\def\hexnumber@#1{\ifcase#1 0\or 1\or 2\or 3\or 4\or 5\or 6\or 7\or 8\or
 9\or A\or B\or C\or D\or E\or F\fi}

\font\tenmsa=msam10
\font\sevenmsa=msam7
\font\fivemsa=msam5
\newfam\msafam
\textfont\msafam=\tenmsa
\scriptfont\msafam=\sevenmsa
\scriptscriptfont\msafam=\fivemsa
\edef\msafam@{\hexnumber@\msafam}
\mathchardef\dabar@"0\msafam@39
\def\dashrightarrow{\mathrel{\dabar@\dabar@\mathchar"0\msafam@4B}}
\def\dashleftarrow{\mathrel{\mathchar"0\msafam@4C\dabar@\dabar@}}

\def\ulcorner{\delimiter"4\msafam@70\msafam@70 }
\def\urcorner{\delimiter"5\msafam@71\msafam@71 }
\def\llcorner{\delimiter"4\msafam@78\msafam@78 }
\def\lrcorner{\delimiter"5\msafam@79\msafam@79 }
\def\yen{{\mathhexbox@\msafam@55}}
\def\checkmark{{\mathhexbox@\msafam@58}}
\def\circledR{{\mathhexbox@\msafam@72}}
\def\maltese{{\mathhexbox@\msafam@7A}}
\def\circledS{{\mathhexbox@\msafam@73}}

\font\tenmsb=msbm10
\font\sevenmsb=msbm7
\font\fivemsb=msbm5
\newfam\msbfam
\textfont\msbfam=\tenmsb
\scriptfont\msbfam=\sevenmsb
\scriptscriptfont\msbfam=\fivemsb
\edef\msbfam@{\hexnumber@\msbfam}
\def\Bbb#1{{\fam\msbfam\relax#1}}
\def\widehat#1{\setbox\z@\hbox{$\m@th#1$}%
 \ifdim\wd\z@>\tw@ em\mathaccent"0\msbfam@5B{#1}%
 \else\mathaccent"0362{#1}\fi}
\def\widetilde#1{\setbox\z@\hbox{$\m@th#1$}%
 \ifdim\wd\z@>\tw@ em\mathaccent"0\msbfam@5D{#1}%
 \else\mathaccent"0365{#1}\fi}
\font\teneufm=eufm10
\font\seveneufm=eufm7
\font\fiveeufm=eufm5
\newfam\eufmfam
\textfont\eufmfam=\teneufm
\scriptfont\eufmfam=\seveneufm
\scriptscriptfont\eufmfam=\fiveeufm

\csname amssym.def\endcsname

\makeatletter
\def\section{\@startsection {section}{1}{\z@}{-3.5ex plus -1ex minus
 -.2ex}{2.3ex plus .2ex}{\large\sc}}
\def\subsection{\@startsection{subsection}{2}{\z@}{-3.25ex plus -1ex minus
 -.2ex}{1.5ex plus .2ex}{\normalsize\sc}}
\makeatother

\makeatletter
\@addtoreset{equation}{section}

\makeatother

\newcommand{\nc}{\newcommand}
\newcommand{\rnc}{\renewcommand}

\nc{\be}{\begin{equation}}
\nc{\ee}{\end{equation}}
\nc{\bea}{\begin{eqnarray}}
\nc{\eea}{\end{eqnarray}}

\nc{\trac}[2]{{\textstyle\frac{#1}{#2}}}

\nc{\ex}[1]{\mbox{e}^{\,\textstyle#1}}

\nc{\CC}{\Bbb{C}}
\nc{\HH}{\Bbb{H}}
\nc{\PP}{\Bbb{P}}
\nc{\RR}{\Bbb{R}}
\nc{\ZZ}{\Bbb{Z}}
\nc{\II}{\Bbb{I}}
\nc{\EE}{\Bbb{E}}
\nc{\SS}{\Bbb{S}}

\rnc{\a}{\alpha}
\nc{\al}{\a^{l}}
\rnc{\d}{\delta}
\nc{\ga}{\gamma}
\nc{\la}{\lambda}
\nc{\lal}{\la_{l}}
\nc{\f}{\phi}
\nc{\fb}{\bar{\phi}}
\nc{\p}{\psi}

\nc{\e}{\eta}
\nc{\eb}{\bar{\eta}}
\rnc{\c}{\chi}
\nc{\eps}{\epsilon}
\rnc{\t}{\theta}
\nc{\tb}{\bar{\theta}}
\nc{\om}{\omega}

\rnc{\P}{\Psi}
\nc{\pl}{\P_{L}}
\nc{\pdr}{\P^{\dag}_{R}}
\nc{\G}{\Gamma}

\nc{\sig}{\sigma}
\nc{\sk}{\sigma_{k}}
\nc{\sa}{\sigma_{a}}
\nc{\Bb}{\bar{B}}

\nc{\symx}{\circledS}

\nc{\Q}{\bar{Q}}
\nc{\C}{{\cal A}/{\cal G}}
\nc{\A}[1]{{\cal A}^{#1}/{\cal G}^{#1}}
\nc{\RC}{{\cal R}_{\C}}
\nc{\RM}{{\cal R}_{\M}}
\nc{\RX}{{\cal R}_{X}}
\nc{\RY}{{\cal R}_{Y}}

\nc{\ad}{\mathop{\mbox{ad}}\nolimits}
\nc{\tr}{\mathop{\mbox{tr}}\nolimits}
\nc{\Tr}{\mathop{\mbox{Tr}}\nolimits}
\nc{\Det}{\mathop{\mbox{Det}}\nolimits}
\rnc{\det}{\mathop{\mbox{det}}\nolimits}
\nc{\rk}{\mathop{\mbox{rk}}\nolimits}
\nc{\diag}{\mbox{diag}}
\nc{\ra}{\rightarrow}
\nc{\Ra}{\Rightarrow}
\nc{\LRa}{\Leftrightarrow}
\nc{\lra}{\leftrightarrow}
\nc{\ot}{\otimes}
\rnc{\ss}{\subset}
\nc{\nul}{\noindent\underline}
\nc{\non}{\nonumber\\}
\rnc{\S}{\Sigma}
\nc{\tp}{2\pi i}
\nc{\del}{\partial}
\nc{\dbar}{\bar{\del}}
\nc{\dx}{\dot{x}}
\nc{\zb}{\bar{z}}

\nc{\mat}[4]{\left(\begin{array}{cc}#1&#2\\#3&#4\end{array}\right)}

\nc{\subs}[1]{{\vspace*{0.5cm}}%
{\noindent\underline{\small\sc #1}}{\addcontentsline{toc}{subsubsection}{#1}}%
{\vspace*{0.3cm}}}

\nc{\chap}[1]{{\clearpage}%
\begin{center}%
{\noindent\underline{\large\sc #1}}{\addcontentsline{toc}{section}{#1}}%
\end{center}%
{\vspace*{0.3cm}}}

\newcommand{\ba}{\begin{eqnarray}}
\newcommand{\ea}{\end{eqnarray}}
\newcommand{\rarw}{\rightarrow}

\begin{document}
\begin{titlepage}

\begin{center}
{\LARGE{\sc $AdS_3 \times{\bf R}$ as a target space}}\\
{\LARGE{\sc for the (2,1) string theory}}\\
\vskip .3in
{\sc M.J. O'Loughlin}
 and
{\sc S. Randjbar-Daemi}\footnote{e-mail: mjol or daemi@ictp.trieste.it}\\
\vspace{.2in}
{\it ICTP\\ Strada Costiera 11, 34014 Trieste\\ Italy}
\end{center}

\begin{abstract}
We study a target space geometry of the form 
$AdS_3 \times {\bf R}$ for the $(2,1)$ heterotic string. This 
target space arises as the near horizon limit of 
a solitonic configuration in $2+2$ dimensions. We investigate
the null isometries of this space and discuss the reduction 
to $1+1$ dimensions of the target space geometry arising from the consistent 
gauging of one of these isometries.
\end{abstract}

\end{titlepage}
\setcounter{footnote}{0}

\section{Introduction}

The $(2,1)$ string theory is a string theory that has an 
ostensibly four dimensional target space of signature $(--++)$.
This four manifold in turn is the world-volume of an extended 
object with a twelve dimensional target space of 
signature $(2,10)$ \cite{KM1KMO}. In addition to 
the equations of motion for the four dimensional geometry,
the $(2,1)$ string theory requires that the four manifold ${\cal M}$
possess an isometry and that in the language of 
the $(2,1)$ sigma model this isometry must be gauged \cite{OV}.
This paper presents a particular non-trivial solution 
to the target space equations of motion that possesses
null isometries. We
discuss the meaning of the gauged isometries of the sigma model
in the context of the action for the $2+2$ dimensional world volume 
(which we shall heretofore refer to as the M-brane). In some respects
the construction reflects recent work in string theory
relating supergravity in anti-de-Sitter space-time to D-brane 
world-volume theories. This similarity arises as the metric
that we consider is the metric on $AdS_3\times {\bf R}$ and
the gauging of the isometry reminds one of the way in which the 
boundary of the $AdS$ space-time is related to the D-brane world-volume. 

In this paper we will restrict ourselves to the null isometries
of ${\cal M}$. The requirement of the null isometry arises as 
the world-sheet of the $(2,1)$ string possess
a U(1) (on the $N=1$ side) that is geometrically realized in the 
target space of the string. The U(1) current must be gauged for 
consistency of the world-sheet theory, and this requires that
the target space possess an isometry \cite{OV}. The requirement of 
freedom from gauge anomalies on the world 
sheet additionally requires that the U(1)
corresponds to a null isometry of the target space. The target space
of this string theory has in addition to the $2+2$ space-time 
co-ordinates eight chiral scalar fields on the left-moving
$N=1$ side. As the $U(1)$ is in the left moving sector of the 
supersymmetry algebra, the null isometry can either lie completely in 
the $2+2$ target space, or it can include a component in one 
of the additional eight directions. From the conformal 
field theory analysis of the $(2,1)$ world sheet one finds that 
in the former case after gauging the isometry one finds a D-string world 
volume as the target space (which is the only case that we will
consider here), and in the latter one finds the D2-brane 
world volume \cite{KM1KMO}. 

The second section of the paper is devoted to an 
overview of the $(2,1)$ string theory, and its relevance to 
recent discussions of M-theory and the unification of 
string theories. In this section we also review the $(2,1)$-sigma model
geometry. This has been discussed in detail in a series of papers
by Hull and Abou-Zeid, and we refer to their papers for a more
complete discussion \cite{hullabou}. 
In the third section we introduce the space-time of interest
that solves the low energy effective field theory of the $(2,1)$
string theory. In the fourth section we look at the null isometries of
this space-time and show that they satisfy the additional consistency 
condition presented in section one. In the final section we discuss the 
mechanics of the reduction process in the context of the 
world-volume action of the M-brane.

\section{An overview of the $(2,1)$ string theory}
The $(2,1)$ string theory provides a potentially unifying framework for
the various known string theories. In this section we firstly present an 
overview of this string theory and in particular the origin of the 
null $U(1)$ current - for more details see the Cargese lectures of 
Martinec \cite{emilcgse}. We then review
the features of the $(2,1)$ sigma model relevant to the 
gauging of this $U(1)$ current.

\subsection{The $(2,1)$ string and M-theory}
The world-sheet theory of the $(2,1)$ string is analogous in 
structure to that of the heterotic $(1,0)$ string theory in 
the manner in which the left and right moving sectors are 
combined. The $N=2$ string theory, has a target space critical 
dimension of four with a complex structure and a signature that 
respects this structure, and therefore for a lorentzian signature
we need a metric of signature $(--++)$.
For the $N=1$ fields we find a standard 
fermionic string living in $9+1$ dimensions. To put the 
left and right moving fields together, thus giving us a  
$2+2$ target space, we need to extend the chiral $9+1$ dimensional
target space (related to the $N=1$ superalgebra) to a $10+2$
dimensional target space, leaving over an internal 8 bosonic 
fields which for reasons of modular invariance must be 
compactified on an $E_8$ lattice, (analogous to the situation for the 
sixteen chiral bosonic fields of the usual heterotic
string theory). 
However, now we of course find that the conformal 
anomaly is no longer saturated and to repair this defect one 
must supplement 
the $N=1$ superconformal algebra by a $U(1)$ current algebra. This 
is in fact fortuitous as the $N=2$ supersymmetry of the right 
moving sector demands that corresponding to the $U(1)$ of the
$N=2$ algebra, the $N=1$ left-moving SUSY algebra needs to have
an anomaly free $U(1)$ current. 

The bosonic fields on the worldsheet are,
\be
x^i_r,\, i = 0\dots 3; x^m_l,\, m=0\dots 11,
\ee
and their fermionic partners are,
\be
\psi^i_r;\, \psi^m_l,
\ee
where the fields with subscript $r(l)$ are right(left) moving 
fields on the worldsheet. Putting the $x^i_{r,l}$ together,
we find fields $x^i$ which are co-ordinates in a four dimensional
target space with signature $(--++)$. The gauged $U(1)$ supercurrent is 
$J_l = v_m\partial x^m_l$, $\Psi_l = v_m\psi^m_l$
and for anomaly freedom of this current we 
need ${\bf v}$ to be a null vector, ${\bf v^2} = 0$ \cite{OV}. 
The effect of this gauged
current is a reduction of the naively $2+2$ dimensional target space
to a $1+1$ or $2+1$ dimensional space depending on whether the
null vector ${\bf v}$ lies entirely in the $2+2$ dimensional part of the 
target, or if it contains a component in one of the time-like directions
and the other component in one of the eight internal dimensions of
the left moving fields. 
For example if ${\bf v}=(1,0,1,0,0,\dots ,0)$ then the target is 
a 1-brane with bosonic co-ordinates $(x^1,x^3)$, fermionic 
partners $(\psi^1, \psi^3)$ and it has the world-volume 
field structure of the D-string of type IIB string theory, 
i.e., it carries eight bosonic co-ordinates and a gauge field
\cite{KM1KMO}. If on 
the other hand ${\bf v}=(1,0,0,0,\dots ,0,1)$ 
then the target is $2+1$ dimensional
and has the action and field content of the D2-brane world-volume theory,
with seven bosonic co-ordinates and a gauge field \cite{KM1KMO}. 
For the details
of this construction, including the explicit vertex operators, and
the conditions arising from the null vector and the Virasoro 
constraints see \cite{emilcgse}.

From the world-sheet theory of the $(2,1)$ string we can also understand
something about the space-time in which the $2+2$ brane (or more properly
its appropriate reduction via ${\bf v}$), is embedded. The vertex operators of
the string theory are the fields of the $2+2$ brane and thus are the 
co-ordinates of the space-time in which it is embedded. In this way one 
discovers the above relationship between 
the choice of ${\bf v}$, D1 or D2 brane
world volumes, and furthermore we learn something about the 
ten dimensional target spaces of these D-branes in the
corresponding  string theories.

One can move on from this point to also find physics of other critical
ten dimensional theories with less supersymmetry, using the basic idea
of the construction of Horava-Witten, wherein the heterotic and type IA
string theories arise from M-theory compactified on $S^1/Z_2$. 
In the context of the $(2,1)$ string, this construction is carried
out by performing an orientifold projection. One then finds,
depending on the choice of the null vector ${\bf v}$, either 
a configuration involving a D-string stretched between two 
D7-branes in a type IIB configuration of the variety that one finds
from F-theory, or one finds a D2-brane stretched between a pair of 
D8-branes. The orientifold projection that is used to construct
these configurations is such that it acts on one time and 
one space direction in the M-brane world-volume. In the 
above, the null direction was chosen to always include the time-like 
direction on which an orientifold projection has been performed. 
Alternatively one can choose the other time-like direction, in which 
case one finds a configuration that has a Dirichlet boundary condition
in a time-like direction. This gives rise to a Euclidean D-brane, 
or E-brane as it is referred to in recent work \cite{ebranes}. 
In the language of that paper, 
we find then that simply a rotation of ${\bf v}$ interchanges
D-branes and E-branes. 

It is important to emphasise the nature of the construction 
of D-branes from the target space of the $(2,1)$ string.
In all of these situations one begins the construction 
with the string theory world-sheet, the actual ten dimensional 
critical string target space is two steps away as it is the 
target space of the M-brane. To fully understand 
the various ten dimensional string theories from this perspective 
really requires a string field theory of the $(2,1)$ string. 
The purpose of the remainder of this paper is to continue the program
initiated in \cite{hullabou, KM2} of studying
directly the M-brane world-volume theory. In particular we 
wish to focus on the issue of the gauging of the $U(1)$ of the string 
theory and its relationship to null isometries of the M-brane
geometry. We introduce a non-trivial target space that possesses
such isometries and show that these isometries satisfy 
the consistency conditions for their 
gauging \cite{hullabou}. We then look at the 
reduction of the action under these isometries and 
the geometry of the D-brane world-volume that remains after the 
gauging.

\subsection{The gauged $(2,1)$ sigma model}

We first review the low energy effective field theory of the $(2,1)$ 
string theory.

The action in $(1,1)$ superspace for the bosonic part of 
the target space is,
\be
S = \frac{1}{4i}\int d^2\sigma d^2\theta (g_{ij}(x) + 
b_{ij}(x)) D_{1+}x^iD_{1-}x^j.
\ee
The $x^i$ are co-ordinates in the target space
${\cal M}$. 

Additionally the action will have $(2,1)$ 
supersymmetry if ${\cal M}$ is even dimensional 
with a complex structure satisfying
\bea
J^i_j J^j_k = -\delta^i_k\non
N^k_{\, ij} = 0
\eea
which is covariantly conserved with respect to the generalised 
connection,
\be
\Gamma^{(+)i}_{jk} = \left\{ \begin{array}{c} i \\ jk \end{array} \right\}
+g^{il}H_{jkl}
\ee
where $H_{ijk} = (db)_{ijk}$ and 
$\left\{ \begin{array}{c} i \\ jk \end{array} \right\}$ 
is the Christoffel connection. 
The complex structure must also be such that,
\be
g_{ij}J^i_kJ^j_l = g_{kl}.
\ee

The action is furthermore conformally invariant and thus a valid
starting point for the corresponding string theory if the 
$\beta$-function equations following from the effective action
\be
S = \int d^Dx\, e^{-2\Phi}\sqrt{-g}(R - \frac{1}{3}H^2 + 4(\nabla\Phi)^2)
\ee
are satisfied and $D=4$\cite{cfmp,hulltown}.

From now on we will use complex co-ordinates on a $2+2$ dimensional
target manifold with co-ordinates, $z^\alpha, \bar{z}^{\bar{\beta}}, \alpha,
\beta = 1,2$. We will also consider only metrics for which the line
element is of the form, $ds^2 = 2g_{\alpha\bar{\beta}}dz^\alpha
d\bar{z}^{\bar{\beta}}$ and the complex
structure is $J_{\alpha\bar{\beta}} = i g_{\alpha\bar{\beta}}$.

The above conditions on the metric, torsion and complex structure
of the target space imply that the geometry is determined by a 
vector field, $k_\alpha$ in the following manner,
\bea
g_{\alpha\bar{\beta}} = \partial_\alpha k_{\bar{\beta}} + 
\partial_{\bar{\beta}} k_{\alpha}\non
b_{\alpha\bar{\beta}} = \partial_\alpha k_{\bar{\beta}} -
\partial_{\bar{\beta}} k_{\alpha}\non
H_{\alpha\beta\bar{\gamma}} = \frac{1}{2}\partial_{\bar{\gamma}}
(\partial_\alpha k_\beta - \partial_\beta k_\alpha)
\eea

It was shown by Hull \cite{whichone} that a sufficient condition 
for the satisfaction of the beta function equations (following from 
the above effective action with $D=4$) is,
\be
\Gamma^{(+)}_i = \Gamma^{(+)i}_{jk}J^k_i = 0,
\ee
and for this solution the dilaton field is given algebraically
by the metric through,
\be
\Phi = - log|detg_{\alpha\bar{\beta}}|.
\ee
It is in fact easy to show that for a complex manifold with 
anti-symmetric tensor field related to the metric as above,
and for which the metric is conformally flat, the vanishing 
of $\Gamma^{(+)i}_{jk}J^k_i$ follows immediately. 
We now turn to a discussion of a particular solution to
these equations which we believe is an illustrative example for
the discussion of theories with two time directions. Notice
that a four manifold that satisfies all of the above conditions
and in addition the conditions that we present
in section 4 following from the gauging of the null isometry \cite{hullabou}
is likely to be quite restricted, and thus one may suspect that there
are no other solutions apart from the one we present here. 

As we see from the above discussion, the equations for the 
target space of the $(2,1)$ string can be reduced to the 
equation $\Gamma_i^{(+)} = 0$, which is then an equation to 
be solved for the vector field $k_i$. This equation follows 
from the effective action \cite{KM2,hullsig}, 
\be
S = \int d^4x\, \sqrt{-det\, g_{\alpha\bar{\beta}}}.
\ee

\section{$AdS_3\times{\bf R}$}

We want to consider a non-compact version of the Hopf manifold,
$S^3 \times S^1$\cite{Hopf}. 
The metric is the product metric on $AdS_3 \times {\bf R}$ with 
a complex structure such that in complex co-ordinates we have,
\bea
ds^2 = {dz_1 d{\bar z}_1 - dz_2 d{\bar z}_2 \over  
z_1 {\bar z}_1 - z_2 {\bar z}_2}\non
= \frac{\eta_{\alpha\bar{\beta}} dz^\alpha dz^{\bar{\beta}}}
{\eta_{\alpha{\bar{\beta}}} z^\alpha z^{\bar{\beta}}}.
\eea
It is easy to see that this is the standard product metric on 
$AdS_3\times {\bf R}$ by noticing that at fixed 
$\rho = \eta_{\alpha{\bar{\beta}}} 
z^\alpha z^{\bar{\beta}}$ we find
the metric on a flat $2+2$ dimensional space-time restricted to
the hyperboloid $\rho =\,$constant (the standard construction of the metric
on $AdS_3$), and then in the radial ($\rho$) direction the 
metric is a flat metric in logarithmic co-ordinates thus the flat 
metric on ${\bf R}$. 

In the euclidean case, this manifold has been discussed in \cite{CHS}, 
as a four dimensional string theory target space with the metric
of $S^3\times{\cal R}$ and a linear dilaton in the flat
direction. The corresponding conformal field theory is then 
an $SU(2)\times U(1)$ WZW model. The $AdS_3$ configuration 
has recently been analyzed as a non-compact 
WZW model in the context of the AdS -- conformal 
field theory correspondence \cite{WZW}. In our case the conformal 
field theory will be $SU(1,1)\times U(1)$ and deserves further 
study in the context of $(2,1)$ string theory.

In fact, this solution is really the near horizon limit
of a solitonic solution that interpolates between 
flat space and a non-trivial configuration as a function of $\rho$. 
This geometry is given by the metric,
\be
ds^2 = (dz_1 d{\bar z}_1 - dz_2 d{\bar z}_2)(1 + {k\over
z_1 {\bar z}_1 - z_2 {\bar z}_2}).
\ee
We now see that the $\rho\rarw 0$ ``near-horizon'' limit or alternatively
the $k\rarw\infty$ large charge limit gives rise
to the geometry of equation (3.1), and therefore we see here a $2+2$
dimensional version of the constructions that originally led Maldacena
to his conjecture relating supergravity to SUSY Yang-Mills theory. 
In \cite{maldaetot}, the near horizon geometry of a solitonic 
D-brane configuration was related to the SUSY Yang-Mills theory that
lives on the D-brane world-volume. In our case we should find
a relationship between the $AdS_3\times\, {\bf R}$ self-dual gravity of 
the $(2,1)$ string target space with null isometry, 
and the D-string world volume theory
that has been shown in \cite{KM2} to arise after 
gauging a null $U(1)$ that lies entirely within ${\cal M}$.

The vector field $k_\alpha$ from which one can derive the metric
and antisymmetric tensor field of (4.1) is,
\be
k_1 = \frac{1}{4z^1} {\rm log} \frac{\rho}{\sqrt{\bar{z}^{\bar{2}}}}
\ee
\be
k_2 = \frac{1}{4z^2} {\rm log} \frac{\rho}{\sqrt{\bar{z}^{\bar{1}}}}
\ee
where the antisymmetric tensor field strength is,
\bea
H_{\alpha\beta\bar{\gamma}} = \frac{1}{2}(g_{\alpha\bar{\gamma},\beta}
- g_{\beta\bar{\gamma},\alpha})\non
= \frac{1}{4\rho^2}(\eta_{\beta\bar{\gamma}}z_\alpha - 
\eta_{\alpha\bar{\gamma}}z_\beta),
\eea
and is identical (up to a factor of $k$ - the charge of the soliton) 
for the $AdS_3\times{\bf R}$ configuration (3.1) and
the interpolating soliton configuration (3.2). As this soliton
carries a non-trivial anti-symmetric tensor field, the 
configuration is appropriately interpreted as the solitonic 
form of the fundamental $(2,1)$ string. Summarising the above, 
we have argued that the near horizon/large charge geometry of the 
solitonic fundamental $(2,1)$ string configuration in the $2+2$ 
dimensional target space of the $(2,1)$ string, may in 
turn be thought of as the $1+1$ dimensional world-volume of
the type IIB D-string embedded in the $1+9$ dimensional space-time
of the IIB string theory. 

As explained at the end of the previous section, this manifold (and
also the interpolating soliton)
will satisfy the equations of motion that arise from the $(2,1)$ string
theory as a consquence of the fact that it is conformally flat and
that the geometry can be encoded entirely in the vector field
$k_\alpha$. From (3.8) we see immediately that the dilaton in 
the M-brane world-volume is proportional to ${\rm log}\, \rho$ and thus 
on each $AdS_3$ represented by an hyperboloid in $R^{2,2}$ the 
dilaton field is a constant, once more as happens in the non-singular 
solitonic D-branes of ten and eleven dimensional supergravities
where Maldacena's conjecture \cite{maldaetot} 
works best - M2-brane, M5-brane, D1$+$D5-brane and D3-brane. 

The isometry group is $SO(2,2) \times {\bf R}$ and is made up 
of the isometries of $AdS_3$ and  translations in the additional 
flat direction. The condition of a gauged $U(1)$ in the world sheet 
theory requires a null isometry in this target space. The invariance
of the complex structure under the isometry addtionally requires that
the isometry  be holomorphic. The additional conditions on 
the null isometry for 
a consistent gauging of the sigma model will be discussed in the 
next section. We will now turn to the Killing vectors of our metric.

The Killing vectors belonging to $SO(2,2)$ in real coordinates $x^i$
are $L_{ij} = x_i\partial_j - x_j\partial_i$, where 
$z^1 = x^1 + ix^2$ and $z^2 = x^3 + ix^4$, where we are for the moment
raising and lowering indices with the flat metric $\eta_{ij}$. 
It is convenient for us to 
consider the following linear combinations of these generators,
corresponding to the isomorphism between $SO(2,2)$ and 
$SU(1,1)\times SU(1,1)$, 
\bea
\xi_{(1)} = \frac{1}{2}(L_{23} - L_{14}) \quad
\eta_{(1)} = \frac{1}{2}(L_{23} + L_{14}) \non
\xi_{(2)} = \frac{1}{2}(L_{31} - L_{24}) \quad
\eta_{(2)} = \frac{1}{2}(L_{31} + L_{24})  \non
\xi_{(3)} = \frac{1}{2}(L_{12} + L_{34}) \quad
\eta_{(3)} = \frac{1}{2}(L_{12} - L_{34}).
\eea
In the complex basis these vectors have the following components,
\bea
\xi_{(1)} =  i(z^2, -z^1)\quad
\eta_{(1)} =  i(\bar{z^2},\bar{z^1})\non
\xi_{(2)} =  (-z^2, -z^1)\quad
\eta_{(2)} =  (-\bar{z^2}, -\bar{z^1})\non
\xi_{(3)} =  i(z^1, -z^2)\quad
\eta_{(3)} =  i(z^1, z^2)
\eea
The remaining $U(1)$ isometry is generated by global scale 
transformations and has components,
\be
S  = (z^1, z^2).
\ee
Of these isometries, clearly $\eta_{(1)}$ and $\eta_{(2)}$ are not 
holomorphic and thus do not preserve the complex structure. 
Of the remaining vectors we have several linear combinations
that are also null. A representative selection of these vectors are
$\xi_{(1,2)} \pm S$ and $\xi_{(1,2)} \pm \xi_{(3)}$.

\section{Null isometries}

The conditions for the gauging of a null isometry $\zeta$ in the target 
space of the $(2,1)$ supersymmetric sigma model was
analyzed in \cite{hullabou}. It is shown in that work that 
the isometry must be holomorphic, that there must be a vector
field $u_\alpha$ solving the equation,
\be
\partial_{[\alpha}u_{\beta]} = \zeta^{\bar{\gamma}}H_{\alpha\beta\bar{\gamma}},
\ee
and there must exist a complex scalar potential $iX + Y$ such that,
\be
\zeta_\alpha  + u_\alpha = \partial_\alpha (iX + Y).
\ee
Note from (4.1) that the solution for $u_\alpha$ has a shift symmetry by the 
gradient of a scalar, and thus up to global considerations 
one can remove the function $Y$ by abosrbing it into $u_\alpha$. 
These fields must satisfy the following additional conditions (see
\cite{hupasp, hullabou} respectively),
\bea
{\cal L}_\zeta  X = 0\non
\zeta^iu_i = 0.
\eea
In words the first of these follows from the equivariance of $u_\alpha$
under the action of the $U(1)$ isometry group  and the
second from the consistency of the WZW term in the $(2,1)$ sigma model.

For our particular solution there are two distinct classes of 
null isometries both of which satisfy the above conditions as
we will now show explicitly. In one class we have the null isometries
of the $AdS_3$ space-time and the other class involves the sum
of an isometry of the $AdS_3$ and a translation in ${\cal R}$. 

The first example that we will consider is the case in 
which the isometry lies entirely within the $AdS_3$ space. The holomorphic 
null vector, $\zeta_{(1)}^\alpha = (\xi_{(3)} - \xi_{(2)})^\alpha$ 
which in complex co-ordinates
is $(iz^1 + z^2, -iz^2 + z^1)$. In this case one finds,
\be
u_\alpha = \frac{1}{2\rho}(i\bar{z}^{\bar{1}} - \bar{z}^{\bar{2}},
i\bar{z}^{\bar{2}} + \bar{z}^{\bar{1}}).
\ee
It is easy to see then that after lowering the indices on $\zeta_{(1)}$ we
have $\zeta_{(1)} + u = 0$ implying $X+iY=$ const. and (4.3) are satisfied.

As a second example consider the null isometry 
$\zeta_{(2)}^\alpha = (S - \xi_{(1)})^\alpha$, 
which in complex co-ordinates is $(z^1 - iz^2, z^2 + iz^1)$.
For this choice of $\zeta_{(2)}$ $u$ is,
\be
u_\alpha = -\frac{i}{2\rho}(\bar{z}^{\bar{2}}, \bar{z}^{\bar{1}})
\ee
and 
\be
(\zeta_{(2)} + u)_\alpha = \frac{1}{2\rho}
(\bar{z}^{\bar{1}},-\bar{z}^{\bar{2}}) 
= \frac{1}{2}\partial_\alpha {\rm log}\,\rho.
\ee
$\zeta + u$ is therefore the gradient of 
\be
iX + Y = \frac{1}{2}{\rm log}\,\rho + {\rm const}.
\ee
Making use of the gauge freedom that allows us to absorb
$Y$ into $u$ we can shift $u$ by $-\partial_\alpha\,Y$ so 
that $\zeta = -u$, $X=$const. and the consistency 
conditions (5.3) are again clearly satisfied.

\section{Reduction along null isometries}

In this section we make a preliminary investigation of the null reduction
from the point of view of the world-volume theory of the M-brane. 
We want to look at the bosonic part of the target space action 
and investigate how the null reduction works. In particular we
will find that the mechanism has some features in common with
the study of singleton fields in $AdS$ spaces
and also appears as an interesting lower dimensional version of
the relationship between gauge theories and supergravity in 
$AdS$ space-times \cite{maldaetot}.
In this paper we will discuss the situation for null vectors
living in the M-brane world volume. 
Null vectors that include some of the internal
left-moving co-ordinates will be left for future work.

We claim that the reduction proceeds by considering the above background,
in the context of the complete $2+2$ dimensional 
Dirac-Born-Infeld type lagrangian derived for this model 
by Kutasov and Martinec \cite{KM2}. Imposing the condition that the fields of 
the model have vanishing Lie derivative along the null 
isometry, we see that this effectively reduces the 
world-volume action from $2+2$ to $1+1$ dimensions. A direct
comparison can then be made between the terms in this
action and the terms in the action for 
the D-string world-volume. The M-brane action is the effective 
action we introduced at the end of section 2, but now we include also
the contributions from the eight target-space scalars \cite{KM2},
\be
S = \int\, d^4x\sqrt{-det(g_{\alpha\bar{\beta}} + \partial_\alpha
\phi^a\partial_{\bar{\beta}}\phi^a)}
\ee
where $a = 1,\dots,8$ and $\alpha = 1,2$. The additional eight 
scalar fields arise as fields in the target space from the $N=1$ 
side of the supersymmetry algebra, just as the $N=0$ sector of the 
$(1,0)$ heterotic string gives rise to target space gauge fields. 
We will expand this action about the above background so that we 
can study the kinetic term only. 

The expansion of the M-brane action to second order in $\phi$ is,
\be
S = \int d^4x(\sqrt{-det(g_{\alpha\bar{\beta}})} +
\frac{(g_{1\bar{1}}A_{2\bar{2}} + g_{2\bar{2}}A_{1\bar{1}} -
g_{1\bar{2}}A_{2\bar{1}} - g_{2\bar{1}}A_{1\bar{2}})}
{2\sqrt{-det(g_{\alpha\bar{\beta}})}} + \cdots)
\ee
where $A_{\alpha\bar{\beta}} = \partial_\alpha\phi^a
\partial_{\bar{\beta}}\phi^a$. In terms of the flat metric 
$\eta_{\alpha\bar{\beta}}$ the kinetic term is proportional to 
$\eta^{\alpha\bar{\beta}}A_{\alpha\bar{\beta}}$. 
Note that this term, quadratic in $\phi$, is conformally invariant,
a property which one would expect for a standard kinetic 
term in {\it two} dimensions.
This is a consequence of the non-standard factor of 
$(-det\, g_{\alpha\bar{\beta}})^{1/2} =  (-det g)^{1/4}$ in the 
M-brane action and is consistent with the fact that the 
dimensional reduction of the M-brane action to $1+1$ dimensions
gives rise to a standard Dirac Born-Infeld action for the type IIB D-string.

To understand more fully what happens in the reduction procedure, 
let us first consider the case for a target space that is 
$2+2$ dimensional flat space. We will consider the kinetic term 
for the eight scalar fields that exist in the target space of the 
$(2,1)$ string, and appear in the square root of the action.
For the flat target space with all other fields turned off
we find the standard kinetic term for the flat metric $\eta_{ij}$ of
signature $(--++)$,
\be
S_{kin} = - \int d^4x\, \partial_i\phi^a\partial_j\phi^a\eta^{ij}
\ee
We consider
the null isometry of ${\cal M}$ corresponding to the vector, 
${\bf v}=(1,0,0,1)$. 
The vanishing of the Lie derivative of the scalar field 
in this direction reduces the kinetic term from a manifestly
four dimensional form to a two dimensional form,
\be
S_{kin} = \int d^4x\, (- \partial_2\phi^a\partial_2\phi^a + 
\partial_3\phi^a\partial_3\phi^a).
\ee

Notice in particular that the null 
direction orthogonal to ${\bf v}$ has also disappeared from 
the action. The meaning of this absence is the new
freedom to add any function of $x_- = x^1 - x^4$ to $\phi$. This 
represents a gauge invariance of the same form
that arises in the discussion of singleton fields in 
$AdS$ space \cite{Frons}. This type of invariance is basically a condition 
that the fields of the theory are chiral in the sense that 
they are functions only of $x_-, x^2$ and $x^3$.
However, we can easily see that the action does not actually 
govern the $x_-$ behaviour of the fields 
and we effectively end up with a theory on a flat $1+1$ dimensional
space-time. 

Returning now to our non-trivial metric we will first change co-ordinates
in the kinetic term using the null Killing vectors.
Consider two pairs of real 
light-cone co-ordinates, $U=x^1+x^4$, $V=x^1-x^4$, $X=x^2+x^3$ and 
$Y=x^2-x^3$ for which the metric takes the form,
\be
ds^2 = \frac{dUdV + dXdY}{UV + XY}
\ee
It turns out to be convenient to consider the 
additional change of variables, $U = R\, cos\,\theta$, 
and $Y = R\, sin\,\theta$.
The Killing vectors that we will consider are then,
\bea
\zeta_{(1)}^i\partial_i = (\xi_{(3)} - \xi_{(2)})^i\partial_i 
= U\partial_X - Y\partial_V,\non
\zeta_{(2)}^i\partial_i = (S - \xi_{(1)})^i\partial_i 
= U\partial_U + Y\partial_Y = R\partial_R.
\eea
In terms of these Killing vectors the kinetic term in the 
world-volume action takes the form,
\be
\int d^4x\,(\frac{\partial_Y\phi^a}{U}
\zeta_{(1)}^i\partial_i\phi^a +
\frac{\partial_V\phi^a}{U}
\zeta_{(2)}^i\partial_i\phi^a).
\ee

When we consider configurations for which $\zeta_{(1)}^i\partial_i\phi = 0$
then we find that the kinetic term has only a single off diagonal 
component,
\be
\frac{R}{U}\partial_R\phi^a\partial_V\phi^a = 
\frac{\partial_R\phi^a\partial_V\phi^a}{cos\,\theta}.
\ee
The vanishing of the Lie derivative of $\phi$ in the $\zeta_{(1)}$ 
direction implies that $cos\,\theta\partial_X\phi - 
sin\,\theta\partial_V\phi = 0$ which implies that $\phi$ 
depends on $X$ and $V$ only in the
combination $y = X\, sin\,\theta + V\,cos\,\theta$. Changing
variables in the surviving part of the kinetic term from
$V$ to $y$, the $cos\,\theta$ in the denominator drops
out and including the factor of $R$ coming from the change to polar 
co-ordinates in the measure we get the final simple diagonal form, 
$R\partial_y\phi^a\partial_R\phi^a$. A further change of variables 
$R=R_0e^t$, results in a flat space kinetic term with $1+1$ dimensional
Lorentz invariance  in almost precise analogy 
with the reduction discussed above for a flat target space. 
We also see that $\phi$ may have, independent of the fixed $\theta$ 
dependence of $y$, an additional undetermined $\theta$ dependence 
analogous to the undetermined $x_-$ depndence of $\phi$ in the flat space
example. 

Considering a general solution to $\zeta_{(2)}^i\partial_i\phi 
= R\partial_R\phi = 0$ the kinetic
term in the action (using also $\partial_Y = sin\,\theta\partial_R + 
{\displaystyle\frac{1}{R}}cos\,\theta\partial_\theta$) is,
\bea
S_{kin} &=& \int dVdXRdRd\theta(\frac{cos\theta\partial_X\phi^a - 
sin\,\theta\partial_V\phi^a}{cos\,\theta})(sin\,\theta\partial_R\phi^a + 
\frac{1}{R}cos\,\theta\partial_\theta\phi^a)\non
&=& \int dVdXdRd\theta
(cos\,\theta\partial_X\phi^a - sin\,\theta\partial_V\phi^a)
\partial_\theta\phi^a.
\eea
Unlike for the gauging of $\zeta_{(1)}$ we cannot further simplify this 
kinetic term since the first term cannot be written as a derivative 
along some co-ordinate direction. On the other hand the metric 
that we deduce from this kinetic term has eigenvalues $0,\,\pm\frac{1}{2}$
and thus we have again ended with a $1+1$ dimensional theory.

In the above examples of ``dimensional reduction'' we have ignored
the behaviour of the field in the 
direction in ${\cal M}$ complementary to the direction of the
null isometry of interest, for example in the discussion of the
reduction for the flat target space we ignored the 
undetermined dependence of $\phi$ on $x_-$. 
To extend the analysis we need to look
at the constrained dynamics of a chiral bosonic field of the 
form $\phi(x_-,x_2,x_3)$ (in the notation used for the 
flat space example) in $2+2$ dimensions, or the similar 
co-ordinate dependence of $\phi$ for the non-trivial metric 
of this paper. 

For the ``gauging'' of the direction $\zeta_{(2)}$ 
a more suggestive approach to the undetermined behaviour of 
$\phi$ in the direction orthogonal to the gauged null direction may
be the following. In our metric the orthogonal
null direction to $\zeta_{(2)}$ is actually, 
$\eta^i\partial_i = V\partial_V + X\partial_X$. 
If the Lie derivatives in this direction are also zero (akin to 
removing the $x_-$ dependence of $\phi$ in the previous paragraph) then 
additionally the Lie derivatives in the direction $\eta + \zeta_{(2)} = S$ 
must vanish, and thus $\phi$ is in particular independent of $\rho$. 
Using a $\phi$ of this form in the kinetic term results in 
(for a fixed value of $\rho$),
\be
\partial_X\phi^a\partial_Y\phi^a(\frac{\rho}{\rho - XY}).
\ee
Furthermore as $\phi$ is independent of $\rho$ 
we may take $\rho\rarw\infty$ to get a $1+1$ dimensional 
kinetic term with a flat Minkowski metric. This is precisely 
the way in which the singleton fields are found as 
$(d-2)+1$ dimensional fields which live on the boundary
at $\rho\rarw\infty$ of the $AdS_d$ space-time, (see \cite{fefr} for 
a recent review). In this case also, the reduction has involved an
additional assumption which is in a similar spirit to the 
holography assumption invoked in the recent work on 
$AdS$ spaces initiated in \cite{maldaetot}.	

The null reduction including fermion fields does not appear 
to be as straightforward since the action is only 
first order in derivatives. Thus the removal of the 
direction complementary (in the above sense) to the direction of
the  null isometry does not appear to happen
in as simple a manner as for the bosonic fields 
described above. In the reduction from $2+2$ to $1+1$ dimensions one does
have the freedom to now restrict the two dimensional chirality of 
the spinors that one obtains from the four dimensional spinors.
This may be sufficient to allow the null reduction to take
us all the way to two dimensions. Related to this subtlety arising 
in the incorporation of fermions in the above construction is
the fact that in the current formulation of the M-brane world-volume
action the correct form for the 
fermionic terms is not currently known \cite{KM2}. 
Turning this discussion around then, it is
possible that formulating the dimensional reduction including
fermions appropriately 
will lead to some new insight into the construction of the 
full supersymmetric low energy effective action for the 
$(2,1)$ string theory.

\subsection*{Acknowledgements}

The authors would like to thank Matthias Blau for useful conversations
during the course of this work and C. Vafa for a useful email exchange.
This work was supported in part by the EC under
the TMR contract ERBFMRX-CT96-0090.

\rnc{\Large}{\normalsize}

\end{document}